\documentclass[nofootinbib,twocolumn,
showpacs,preprintnumbers,amsmath,amssymb]{revtex4}
\usepackage{latexsym,graphicx,amsfonts,amssymb,array,calc,tabularx,dsfont,amsthm,bm,mathrsfs,color}
\usepackage{epsfig, psfrag, graphicx,color,dsfont}

\newcommand{\be}{\begin{equation}}
\newcommand{\ee}{\end{equation}}
\newcommand{\ben}{\begin{eqnarray*}}
\newcommand{\een}{\end{eqnarray*}}
\newcommand{\bea}{\begin{eqnarray}}
\newcommand{\eea}{\end{eqnarray}}
\newcommand{\bdm}{\begin{displaymath}}
\newcommand{\edm}{\end{displaymath}}
\newcommand{\ba}{\begin{align}}
\newcommand{\ea}{\end{align}}
\newcommand{\lb}{\label}

\renewcommand{\exp}{\operatorname{exp}\!}
\renewcommand{\cosh}{\operatorname{cosh}\!}
\renewcommand{\sinh}{\operatorname{sinh}\!}
\renewcommand{\cos}{\operatorname{cos}\!}
\renewcommand{\sin}{\operatorname{sin}\!}

\newcommand\R{\mathds{R}}
\newcommand\Nmu{\mathcal{N}_{\mu}(c)}
\newcommand\Eref{Eq.~\eqref}

\begin{document}

\title{Quintessence and (Anti-)Chaplygin Gas in Loop Quantum Cosmology}

\author{Raphael Lamon\footnote{e-mail address: {\tt raphael.lamon@uni-ulm.de}} and Andreas J. W\"ohr\footnote{e-mail address: {\tt andreas.woehr@uni-ulm.de}}}
\affiliation{Institut f\"ur Theoretische Physik, Universit\"{a}t Ulm, Albert-Einstein-Allee 11,
89069 Ulm, Germany}

\begin{abstract}
The concordance model of cosmology contains several unknown components such as dark matter and dark energy. Many proposals have been made to describe them by choosing an appropriate potential for a scalar field. We study four models in the realm of loop quantum cosmology: the Chaplygin gas, an inflationary and radiationlike potential, quintessence and an anti-Chaplygin gas. For the latter we show that all trajectories start and end with a type II singularity and, depending on the initial value, may go through a bounce. On the other hand the evolution under the influence of the first three scalar fields behaves classically at times far away from the big bang singularity and bounces as the energy density approaches the critical density.
\end{abstract}

\pacs{04.60.Pp,98.80.Cq}

\maketitle

\section{Introduction}\label{sec:intro}
It is generally believed that our Universe started with an inflationary phase followed by a radiation and matter dominated era. However, classical cosmology is not able to tackle the problem of the initial conditions of the universe. One of the possible solutions to this problem is that our expanding Universe was preceeded by a contracting phase. But powerful singularity theorems based on classical general relativity (GR) forbid such a behavior unless one assumes a form of matter that violates the positive energy conditions or modified versions of gravity (see e.g. \cite{Abdalla:05,Nojiri:08}). On the other hand it is believed that quantum gravity should solve this problem by generating ideal conditions for the genesis of our universe. Several proposals such as the pre-big bang string cosmology\cite{Gasperini:03} and the ekpyrotic/cyclic models\cite{Khoury:01,Steinhardt:02} modify dynamics with (perturbative) quantum gravitational effects but have so far had limited viability.

A generic nonsingular evolution through the big bang can only be achieved if nonperturbative effects of quantum gravity are incorporated. One of the leading nonperturbative background independent approach is loop quantum gravity (LQG)\cite{Ashtekar:04,Rovelli:04,Thiemann:07}. One of the main predictions of LQG is that the underlying geometry is discrete. The application of the quantization methods of LQG to homogeneous spacetimes results in what is known as loop quantum cosmology (LQC) \cite{Bojowald:08,Bojowald:00,Bojowald:00:2,Bojowald:00:3,Bojowald:02,Ashtekar:03,Bojowald:03}. The results of LQC not only provide new insights into the quantum structure of spacetime near the big bang singularity but also remove this singularity by extending the time evolution to negative times. It has been rigorously shown \cite{Ashtekar:06,Ashtekar:06:2,Ashtekar:08} that the evolution of contracting semiclassical universes passes through the quantum regime and expands to semiclassical universes. This nonsingular bounce stems from the fact that the dynamics in LQC is governed by a discrete quantum difference equation in quantum geometry. On the other, it can be shown \cite{Ashtekar:06,Ashtekar:06:2,Singh:05,Vandersloot:05} that an effective Hamiltonian on a continuum spacetime can be found which approximates well the quantum dynamics (for a Wheeler-DeWitt analog see \cite{Nelson:08}). The modification arising from nonperturbative effects to the classical Friedmann equation includes a term $\rho^2/\rho_{\mathrm{crit}}$, where $\rho$ is the energy density and $\rho_{\mathrm{crit}}$ denotes the critical density of the order of magnitude of the Planck density. Since this term is negative the evolution bounces whenever the energy density reaches a density close to the Planck density.

The viability of the bounces for more general matter sources has been studied in e.g. \cite{Singh:06,Nelson:07,Nelson:07:2, Cailleteau:08}, where it was shown that the behavior of solutions with inflationary and negative potentials are nonsingular, respectively, where solutions of exponential potentials are analyzed. Moreover it was shown that for negative potentials the inner boundary also appears, corresponding to the classical recollapse, which leads to solutions having cyclic behavior. In \cite{Cailleteau:09,Chiou:07} the authors studied the role of LQC effects in the Ekpyrotic/Cyclic model in Bianchi type I models and showed that the universe undergoes multiple small bounces and the anisotropic shear remains bounded throughout the evolution. 

In this work we are interested in potentials which play a major role in classical cosmology. We first introduce the effective dynamics in LQC in Sec.~\ref{sec:effdyn}. In Sec.~\ref{sec:Singularities} we give a short overview of conditions for singularities occuring in Friedmann-Robertson-Walker (FRW) cosmologies. In Sec.~\ref{sec:chaplygin} we study the Chaplygin gas and in Sec.~\ref{sec:inflrad} we study the robustness of the bounce for a scalar field which has the property of being inflationary at small times and radiationlike at later times \cite{Bouhmadi:09:2,Bouhmadi:09,Bouhmadi:09:3}. Sec.~\ref{sec:antich} is devoted to the anti-Chaplygin gas and Sec.~\ref{sec:quintessence} to a quintessence model which models dark energy. We conclude with Sec.~\ref{sec:conclusion}.

\section{Effective Dynamics in LQC}\label{sec:effdyn}

LQG is a canonical quantization of gravity based upon Ashtekar connection variables. The phase space of classical GR in LQG is spanned by a SU(2) connection $A^i_a$ and a densitized triad $E_i^a$ on a 3-manifold ${\cal M}$, which are two conjugate variables encoding curvature and spatial geometry, respectively. Likewise, LQC is a canonical quantization of homogenous spacetimes such that the phase space structure is simplified, i.e., the connection is determined by a single quantity labeled $c$ and likewise the triad is determined by a parameter $p$. For the spatially flat model of cosmology, the new variables are related to the metric components of the (FRW) universe through
\be 
\label{c,p}
c = \gamma \dot{a} , \qquad p = a^2 \ ,  
\ee
where $\gamma$ is the Barbero-Immirzi parameter which is set to be $\gamma \approx 0.2375$ by the black hole entropy
considerations \cite{ABC:98}. Classically in terms of the connection-triad variables the Hamiltonian constraint is given by
\be \lb{Hclass}
{\cal H}_{\mathrm{cl}} = - \frac{3 \sqrt{p}}{\kappa \gamma^2} c^2 + {\cal H}_{\mathrm{M}}
\ee
with $\kappa = 8 \pi G$ (where $G$ is Newton's gravitational constant) and ${\cal H}_{\mathrm{M}}$ being the matter Hamiltonian. The complete equations of motion are derived from Hamilton's equations $\dot{x} = \{ x , {\cal H}_{\mathrm{cl}} \}$ for any phase space variable $x$, and by enforcing that ${\cal H}_{\mathrm{cl}}$ should vanish. The variables $c$ and $p$ are canonically conjugate with Poisson bracket $\{c,p\} = \gamma \kappa / 3$. The classical Friedmann equations are obtained through a substitution of these relations into the Hamiltonian constraint (\ref{Hclass}).

The basic variables of LQC are the component of the densitized triad and the holonomies along straight edges: $h_i(\mu) = \exp \,(\mu c\tau_i) = \cos\,(\mu c / 2) + 2 \sin\,(\mu c / 2 )\, \tau_i$, where $\tau_i$ is related to the Pauli spin matrices through $\tau_i = - i \sigma_i/2$. The dimensionsless parameter $\mu$ represents the physical length of the edge and is arbitrary. As such, the almost periodic functions $\exp\,(i\mu c/2)=:\Nmu$, $\mu\in\R$, can be chosen to be the elementary variables of LQC. The operator $\hat p$ corresponding to the component of the densitized triad acts by differentiation and is diagonalized by $\Nmu$. In a canonical setting, the dynamics is implemented completely by the Hamiltonian constraint. Upon quantization, the Hamiltonian constraint is promoted to an operator using Thiemann's trick \cite{Thiemann:98}.

It has been shown that the underlying dynamics in LQC is governed by a discrete difference equation in eigenvalues $V_{\mu}$ of the volume operator $\hat V$ in quantum geometry (see e.g. \cite{Bojowald:08}). However, an effective Hamiltonian description on an continuum spacetime can be constructed using semiclassical states, which has been shown to very well approximate the quantum dynamics \cite{Ashtekar:06,Ashtekar:06:2}. The analysis of the quantum Hamiltonian using semiclassical states belonging to the physical Hilbert space reveals that a backward evolution of our expanding phase of the universe leads to a bouncing solution into a contracting branch \cite{Ashtekar:06:3}. The expectation values of the Dirac observables allows us to investigate to quantify the difference between the quantum and classical dynamics. It turns out that quantum geometric effects become dominant only when the energy density $\rho$ of the universe is of the order of the critical density $\rho_{\mathrm{crit}}$ \cite{Ashtekar:06:2,A:06} and classical general relativity is a good approximation to the quantum dynamics when $\rho\ll \rho_{\mathrm{crit}}$. 
The effective equations for the modified Friedmann dynamics can be derived from the effective Hamiltonian constraint with loop quantum modifications. The effective Hamiltonian constraint, to leading order, is given by \cite{Singh:05}
\be
{\cal H}_{\mathrm{eff}} = - \frac {3}{\kappa \gamma^2 \bar \mu^2} a \sin^2(\bar \mu c) + {\cal H}_{\mathrm{M}} \ .
\ee
where $\bar \mu = \sqrt{3\sqrt{3}/{2|\mu|}}$ \cite{Ashtekar:06:2}.

In this work we will be mainly interested in the matter Hamiltonians corresponding to a
massive scalar field $\phi$ with momentum $\Pi_\phi$ and potential $V(\phi)$:
\be 
\lb{hmatter}
{\cal H}_{\mathrm{M}} = \frac{1}{2} \frac{\Pi_\phi^2}{p^{3/2}} + p^{3/2}  V(\phi) \ .
\ee
The energy density and pressure of the scalar field are given by
\be 
\lb{edp}
\rho = \rho_\phi = \frac{1}{2} \, \dot \phi^2 + V(\phi), \qquad  p_\phi = \frac{1}{2} \, \dot \phi^2 - V(\phi)\ .
\ee
The scalar field satisfies the stress-energy conservation law:
\be
\lb{EC} 
\dot \rho_\phi + 3 \frac{\dot a}{a} (\rho_\phi + p_\phi) = 0 \ .
\ee

The modified Friedmann equation for $\dot p$ is obtained with Hamilton's equations
\be
\lb{dotp}
\dot{p} = \{p,{\cal H}_{\mathrm{eff}}\} = - \frac{\gamma \kappa}{3} \frac{\partial}{\partial c}{\cal H}_{\mathrm{eff}}
 = \frac{2a}{\gamma \bar \mu} \sin\left(\bar {\mu} c\right)  \cos \left(\bar {\mu} c\right)
\ee
which on using Eq.~\eqref{c,p} implies that the rate of change of the scale factor
is given by
\be
\lb{dota} 
\dot {a} = \frac{1}{\gamma \bar \mu} \sin\left(\bar {\mu} c\right)  \cos \left(\bar {\mu} c\right) \ . 
\ee
Furthermore, the vanishing of the Hamiltonian constraint implies
\be
\lb{sin}
\sin^2\left(\bar {\mu} c\right)  = \frac{\kappa \gamma^2 \bar \mu^2}{3a} {\cal H}_{\mathrm{M}}\ .
\ee
Combining Eqs.\eqref{dota} and \eqref{sin} yields the effective Friedmann equation for the Hubble rate $H=\dot{a}/a$
\be 
\lb{FE1}
H^2 = \frac{\kappa}{3} \rho \left(1 - \frac{\rho}{\rho_{\mathrm{crit}}} \right)  \ .
\ee 
with the critical density given by 
\be
\lb{ecr}
\rho_{\mathrm{crit}} = \frac{\sqrt{3}}{16 \pi^2 \gamma^3}\rho_{\mathrm{pl}} \ , 
\ee
where $\rho_{\mathrm{pl}} = 1/(\hbar G^2)$ is the Planck density. The modified Friedmann equations provides an effective description for LQC which approximates the underlying quantum dynamics very well. The $\rho^2$-modification of the Friedmann equation arises from nonperturbative quantum geometric effects. Since the modified term is negative definite, the Hubble parameter vanishes when $\rho = \rho_{\mathrm{crit}}$ and the universe experiences a turn-around in the scale factor. For $\rho \ll\rho_{\mathrm{crit}}$, the modifications become negligible such that the standard Friedmann equations are recovered. In addition, it should be noted that $\rho_{\mathrm{crit}}$ is of quantum origin since $\rho_{\mathrm{pl}} \propto 1/\hbar$.

The modification arising in \Eref{FE1} constrains the Hubble parameter and the energy density to be bounded from below and above:
\be
\lb{HB}
H\in[-\sqrt{\frac{\kappa\rho_{\mathrm{crit}}}{12}},\sqrt{\frac{\kappa\rho_{\mathrm{crit}}}{12}}] \quad \text{and}\quad \rho\in[0,\rho_{\mathrm{crit}}]\ .
\ee

Moreover, using the conservation law \eqref{edp}, its time derivative can be cast into
\be\label{Hdot}
\dot H=-\frac{\kappa}{2}(\rho+p_{\phi})\left(1-\frac{2\rho}{\rho_{\mathrm{crit}}}\right).\ee
The Hubble parameter can be expressed in terms of the scalar field such that
\be\label{phiddot}
\ddot\phi=-\frac{\partial V}{\partial\phi}\mp3\dot\phi\left[\frac{\kappa}{3}\rho\left(1-\frac{\rho}{\rho_{\mathrm{crit}}}\right)\right]^{\frac{1}{2}},\ee
where expansion corresponds to the upper sign and contraction to the lower sign.

Apart from the case of a free scalar field it is difficult to find analytical solutions. For this reason we will draw phase portraits showing the qualitative behavior of the numerical solutions. Using the Hamiltonian~\eqref{Hclass} the four-dimensional parameter space $(c(t),p(t),\phi(t),\dot\phi(t))$ can be reduced by one unit by expressing one of these variables by the other three. Following \cite{Singh:06} we will display a phase portrait consisting of the variables $\phi$ and $\dot\phi$. The quantum turn-arounds will be represented as solid lines showing a boundary for the solutions. Once a trajectory reaches such a boundary the sign of the Hubble rate changes, thus indicating a turnaround. Since our potentials are complicated we will also show the phase portraits illustrating the time evolution of the kinetic and potential energy.

\section{Cosmological Singularities}\label{sec:Singularities}
Over the last few years, the zoo of cosmological singularities has become considerably more extensive. Beside the traditional singularities known as big bang an big crunch, there also exist the big rips and sudden singularities. In this section we provide a catalog of relevant singularities in FRW cosmologies \cite{Nojiri:05, Lazkoz:06, Visser:05,Singh:09}. All singularities are classified by means of both the kinematic (scale factor $a$) and the dynamical (energy density $\rho$ and pressure $p$) behavior .

\textit{Big bang and big crunch}: the most basic of the cosmological milestones are big bangs and big crunches, for which the scale factor $a(t)\rightarrow0$ at some finite time as we move to the past or future. Also both the energy density $\rho$ and curvature invariants diverge. Dynamically, the Null Energy Condition (NEC hereafter), $\rho+p\geq0$, is always satisfied.
\newline\textit{Big rip or type I singularity}: a big rip is said to occur if $a(t)\rightarrow\infty$ at some finite time \cite{Caldwell:02,Caldwell:03}. This is accompanied with a divergence of the energy density, pressure and curvature invariants. These singularities always violate NEC and all other energy conditions such as the WEC (NEC \& $\rho\geq0$), SEC (NEC \& $\rho+3p\geq0$) and DEC ($\rho\pm p\geq0$) \cite{Visser:05}. The type I singularity emerges for phantom-like equations of state: $w<-1$.
\newline\textit{Sudden or type II singularity}: this extreme event is characterized by a finite value of the energy density but an associated divergence of pressure at finite value of the scale factor. Due to the divergence of pressure, the Ricci curvature scaler $R$ diverges. 
\newline\textit{Type III singularity}: like type II singularities, but the energy density and pressure diverge, causing a blow up of curvature invariants at some finite time. The type III appears for the model with $p=-\rho-A\rho^{\alpha}$ \cite{Odintsov:04,Stefancic:05}. 
\newline\textit{Type IV singularity}: higher-order time derivative of the scale factor $a$ diverge at finite time, while the scale factor itself remains finite. None of the energy density or pressure blows up in this case. The type IV singularity appears in models characterized by $p=-\rho-f(\rho)$, where $f(\rho)$ can be an arbitrary function\cite{Nojiri:05}.
\section{Chaplygin Gas}\label{sec:chaplygin}
The Chaplygin gas was introduced to cosmology in \cite{Kamenshchik:01} to describe the transition from a universe filled with dust-like matter to an exponentially expanding universe. This gas is a perfect fluid which has the following equation of state:
\be\label{CHequofstate} p=-\frac{A}{\rho},\ee
where $A$ is a positive constant. Energy conservation requires that
$$\rho=\sqrt{A+\frac{B}{a^6}},$$
where $B$ is an integration constant. For positive $B$ and small $a$ we get a universe dominated by dust-like matter:
$$\rho\sim\frac{\sqrt{B}}{a^3},\quad a^6\ll\frac{B}{A}.$$
For large $a$ it turns out that the universe is of the de~Sitter type with a cosmological constant $\sqrt{A}$:
$$\rho\sim\sqrt{A}, \quad a^6\gg\frac{B}{A}.$$
The potential corresponding to this equation of state is given by \cite{Kamenshchik:01}
\begin{align}\label{CHpotential} V_{\mathrm{CH}}(\phi)&=\frac{2a^6\left(A+\frac{B}{a^6}\right)-B}{2a^6\sqrt{A+\frac{B}{a^6}}}\nonumber\\
&=\frac{1}{2}\sqrt{A}\left(\cosh\sqrt{3\kappa}\phi+\frac{1}{\cosh\sqrt{3\kappa}\phi}\right).
\end{align}
A generalization of this gas has been introduced in \cite{Bento:02}, where the equation of state is given by
$$p=-\frac{A}{\rho^{\alpha}},$$
where $\alpha$ is a positive constant. The requirement that the sound velocity not exceed the speed of light yields to the bound $0<\alpha\leq 1$. The potential for the scalar field corresponding to this equation of state reads
$$V_{\mathrm{GCH}}(\phi)=V_0\left[\cosh\left(\sqrt{\kappa}\beta\phi\right)^{\frac{2}{\alpha+1}}+\cosh\left(\sqrt{\kappa}\beta\phi\right)^{\frac{-2}{\alpha+1}}\right],$$
where $\beta=\sqrt{3(\alpha+1)/2}$. Leaving both $A$ and $\alpha$ free, the latest cosmological and astrophysical constrain these parameters to the following $1\sigma$ confidence level \cite{Lu:09}
$$\alpha=-0.09^{+0.15}_{-0.12} \quad\mathrm{and}\quad A_s=0.73^{+0.06}_{-0.09},$$
where $A_s=A/(A+B)$ and $B$ is an integration constant. However, we checked numerically the influence of $\alpha$ and it turns out that different values of $\alpha$ do not change the behavior of the bounce. The reason is that, since $a\rightarrow0$ implies $|\phi|\rightarrow\infty$, we have
$$V_{\mathrm{GCH}}\sim V_0\exp\left(\sqrt{\kappa}\beta|\phi|\right)^{\frac{2}{\alpha+1}}.$$
Thus, different values in the parameter space $(A,\alpha)$ only change the magnitude of the potential and not its shape. From now on we only consider the case $\alpha=1$.

\begin{figure}[!ht]
 \begin{center}
 \includegraphics[width=7cm]{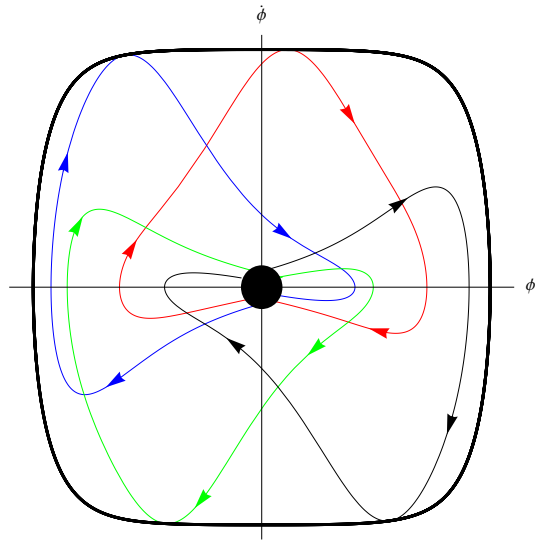}
\caption{Phase portrait for the Chaplygin gas with potential~\eqref{CHpotential} for $\sqrt{A}=10^{-3}$. The thick line shows the boundary indicating turn-arounds, the thin lines show the solutions of \Eref{phiddot} for different initial data.}\label{fig:CHOK_paramplot}
\end{center}
\end{figure}

\begin{figure}[!ht]
 \begin{center}
 \includegraphics[width=7.5cm]{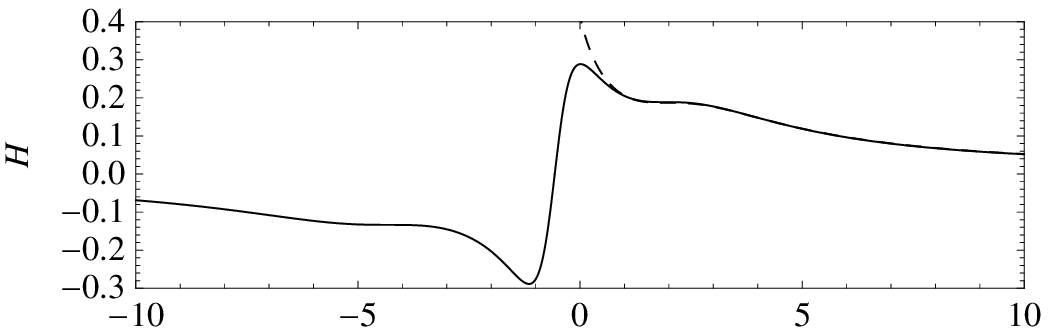}
\includegraphics[width=7.5cm]{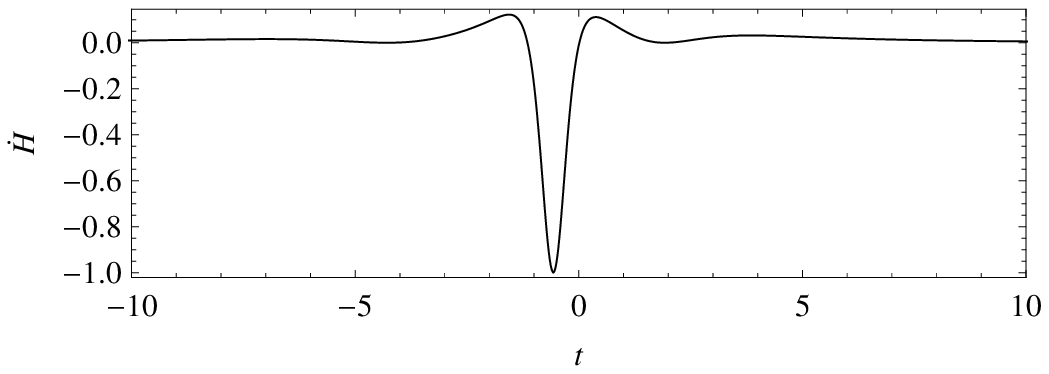}
\caption{Graphs of the Hubble rate and its time derivative for the Chaplygin gas as a function of time. The dashed line represents the classical solution and the solid lines the solution from LQC. The bounce occurs at $t=0.52$, where the Hubble rate changes sign.}\label{fig:CHOK_cl_vs_LQC}
\end{center}
\end{figure}

\begin{figure}[!ht]
 \begin{center}
 \includegraphics[width=7.5cm]{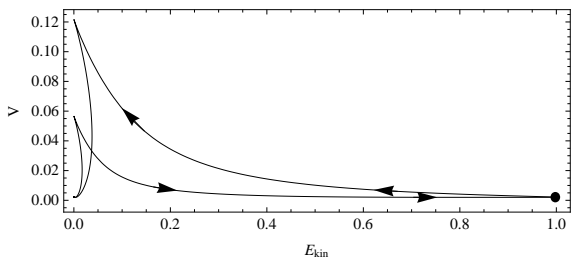}
\caption{Phase portrait of the kinetic energy $E_{\mathrm{kin}}=\dot\phi^2/2$ and the potential energy $V_{\mathrm{CH}}(\phi)$. The dot corresponds to the bounce. The kinetic energy vanishes at $t=-4.3$ and $t=1.9$.}\label{fig:CHOK_kinV}
\end{center}
\end{figure}

Fig.~\ref{fig:CHOK_paramplot} shows the phase portrait of the variables ($\phi(t)$,$\dot\phi(t)$) for four different initial values. All trajectories start from the point $(0,0)$ for $t\rightarrow-\infty$ where the energy density vanishes. 
Also, the Hubble rate is negative but close to zero (see Fig.~\ref{fig:CHOK_cl_vs_LQC}). As can be seen from Fig.~\ref{fig:CHOK_kinV} the potential energy is the dominant contribution for times far away from the bounce. As the energy density starts to grow the trajectories depart from $(0,0)$. At time $t=-4.3$ the kinetic enery vanishes and the potential energy reaches a local maximum, which can also be seen from the plateau in the Hubble rate. The evolution then reaches the point at the boundary where the universe bounces. The energy density is highest at this point (denoted by the dot in Fig.~\ref{fig:CHOK_kinV}) and the dominant contribution comes from the kinetic energy. Moreover, as a manifestation of the bounce the sign of the Hubble rate changes and becomes positive. Then the evolution reaches a second plateau at time $t=1.9$ where the kinetic energy vanishes and the potential has its global maximum. For $t\rightarrow\infty$ all trajectories go to the point $(0,0)$ of the phase portrait and the Hubble rate decreases with the same rate as radiation, i.e. $H\sim (2t)^{-1}$.

\section{Inflationary and Radiationlike potential}\label{sec:inflrad}

\begin{figure}[!ht]
 \begin{center}
 \includegraphics[width=7cm]{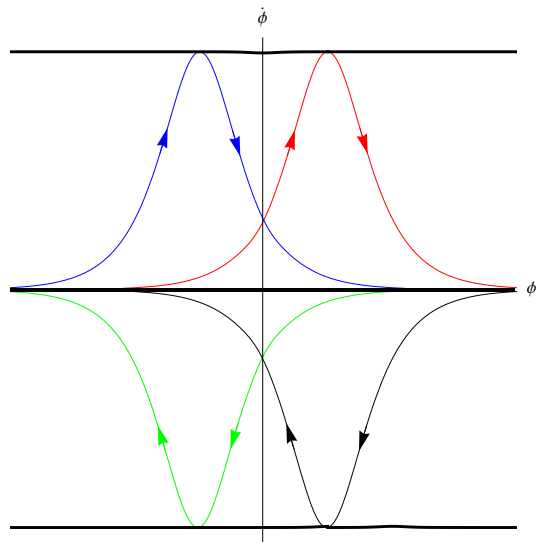}
\caption{Phase portrait for a scalar field with potential~\eqref{potinflrad} for $V_0=10^{-2}$. The thick lines represents the boundaries where the solution of \Eref{phiddot} undergo a bounce.}\label{fig:BouncePlotsMiriam_paramplot}
\end{center}
\end{figure}

In this section we study a scalar field with a scalar filed potential which can be seen as a modification to the Chaplygin gas. The general motivation behind this modification is the fact that the potential can be used to model both radiation and inflation (see also \cite{Zhang:07,Zhang:09}). 
Let us consider the energy density\cite{Bouhmadi:09:2,Bouhmadi:09,Bouhmadi:09:3}:
\be\label{rhoinflrad} \rho=\left(A+\frac{B}{a^{4(1+\alpha)}}\right)^{\frac{1}{1+\alpha}},\quad 1+\alpha<0,\ee
where $A$, $B$ and $\alpha$ are constants. For early times the energy density is inflationary:
$$\rho\sim A^{\frac{1}{1+\alpha}}, \quad A\gg B/a^{4(1+\alpha)},$$
and for late times radiationlike:
$$\rho\sim\frac{1}{a^4},\quad B/a^{4(1+\alpha)}\gg A.$$
Such a behavior can be modeled by a scalar field with the following potential \cite{Bouhmadi:09:2,Bouhmadi:09,Bouhmadi:09:3}:
\begin{align}\label{potinflrad} V_{\mathrm{IR}}(\phi)=\frac{V_0}{3}\biggl[&\cosh^{\frac{2}{1+\alpha}}(-k(1+\alpha)\phi)\nonumber\\
&+2\cosh^{-\frac{2\alpha}{1+\alpha}}(-k(1+\alpha)\phi)\biggr].\end{align}
This potential shares many similarities with the Chaplygin potential~\eqref{CHpotential}. On the other hand, while the potential energy of the Chaplygin gas is the dominant contribution to the energy density at late times and just after (respectively before) the bounce (see Fig.~\ref{fig:CHOK_kinV}), $V_{\mathrm{IR}}(\phi)$ is always at least one order of magnitude smaller than the kinetic term. So, instead of going to zero as can be seen in Fig.~\ref{fig:CHOK_paramplot}, $\phi\rightarrow\pm\infty$ as $t\rightarrow\pm\infty$ (see Fig.~\ref{fig:BouncePlotsMiriam_paramplot}). Moreover, the Hubble parameter $H$ decreases at a radiationlike rate, i.e. $H\sim 1/(2t)$. As the evolution approaches the classical singularity both the kinetic and potential energy densities increase and approach the critical density. This is when LQC modifications come into play such that the evolution goes through a bounce. This point is reached when the evolutions represented by the thin lines in Fig.~\ref{fig:BouncePlotsMiriam_paramplot} touch the boundary shown as thick lines.

\section{Anti-Chaplygin Gas}\label{sec:antich}
The anti-Chaplygin gas was introduced in the context of cosmology in \cite{Kamenshchik:07}. The pecularity of this gas is that its equation of state is given by
$$p=\frac{A}{\rho},$$
which can be modeled by a scalar field with potential
\be\label{potACH}V_{\mathrm{ACH}}(\phi)=V_0\left(\sinh\,(\sqrt{3\kappa}|\phi|)-\sinh^{-1}(\sqrt{3\kappa}|\phi|)\right) \ .
\ee
In a cosmological context such a gas leads to a big brake singularity caused by the divergence of of higher derivatives of the Hubble rate. This singularity occurs at a finite value of the scale factor where the Hubble rate vanishes. Since the second time derivative of the scale factor diverges the Ricci scalar $R$ also diverges. Moreover, while the energy density remains finite the pressure diverges. Such singularities are called sudden of type II singularity \cite{Singh:09}.

\begin{figure}[!ht]
 \begin{center}
 \includegraphics[width=7cm]{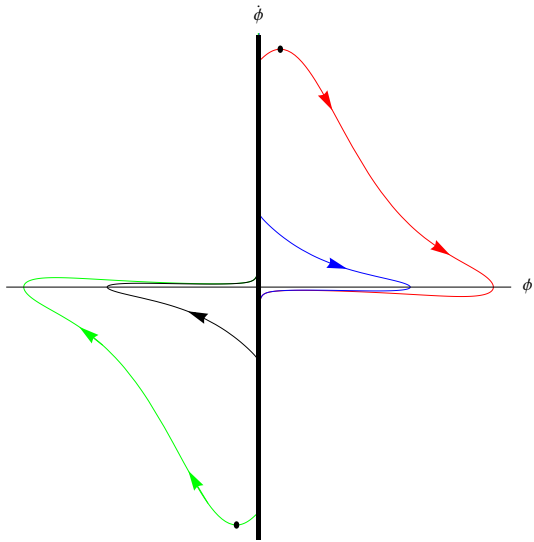}
\caption{Phase portrait for the anti-Chaplygin gas with potential~\eqref{potACH} for $V_0=10^{-4}$. The thick line represents the singularity $V_{\mathrm{ACH}}(\phi)\rightarrow-\infty$, the thine lines the solution of \Eref{phiddot} and the dots the bounce. The evolution starts and ends at a type II singularity.}\label{fig:AntiCHOK_paramplot}
\end{center}
\end{figure}

The energy density is near zero when the type II singularity occurs such that the modifications arising from LQC are not able to avoid this divergence. This can be seen in Fig.~\ref{fig:AntiCHOK_paramplot} where every solution converges toward $\phi=0$. The reason why the energy density does not diverge is because the kinetic term in \Eref{edp} cancels the divergence from the potential. However, the pressure diverges because the potential is unbounded from below. On the other hand LQC is able to remove the Big Bang singularity occuring at times represented by dots in Fig.~\ref{fig:AntiCHOK_paramplot}. From a backward evolution perspective, not every solution goes through a bounce because, depending on the initial value, the third singularity is reached. This singularity is also of type II because the energy density is finite but the pressure diverges. As before, LQC is not able to resolve it and the evolution stops.

In sum, the evolution of a universe filled with an anti-Chaplygin gas starts and stops at a type II singularity when the point $\phi=0$ is reached. Depending on the initial value, it may go through a bounce.

\section{Quintessence}\label{sec:quintessence}
Recent observations of the anisotropy of the cosmic mircowave background (CMB) \cite{Hinshaw:09} together with the power spectrum of the large scale structure (LSS) \cite{Chang:08} and the magnitude-redshift relation of the supernovae Ia \cite{Riess:07,Lampeitl:09} all indicate that the current mean energy density $\rho_{tot}$ of the Universe consists not only of radiation, baryonic and dark matter, but also of a dominant component of negative pressure form which is called dark energy.
An explanation for the missing energy is quintessence where the dark energy density is identified with the energy density $\rho_{\phi}$ (associated with a negative pressure $p_{\phi}$) arising from a scalar (quintessence) field $\phi$.
It is possible to construct scalar field potentials $V$ which lead to a constant equation of state $w_{\phi}=p_{\phi}/\rho_{\phi}$. The form of such potentials depends on the energy components of the model. Potentials with three components (radiation, matter and qiuntessence) can only be modeled for special values of $w_{\phi}$. The exact quintessence potential for $w_{\phi}=-\frac{1}{3}$ reads \cite{Steiner:02}
\be
\lb{SteinerPot}
V_{\mathrm{Q}}(\phi)=\frac{V_0}{\left[\eta\sinh\,(B\phi)+\cosh\,(B\phi)-1\right]^2} \ , 
\ee
where the potential strength $V_0$, respectively $B$ and $\eta$ are given by
\be
\lb{V0}
V_0=\frac{8}{3}\frac{\Omega_{\phi}\Omega^2}{\Omega^2_{\mathrm{m}}}\rho_0,\quad  B=\frac{2\sqrt{\pi}}{m_{\mathrm{pl}}}\sqrt{\frac{\Omega}{\Omega_{\phi}}},\quad \eta=2\frac{\sqrt{\Omega\Omega_{\mathrm{rad}}}}{\Omega_{\mathrm{m}}}
\ee
and $\Omega=1-\Omega_{\mathrm{rad}}-\Omega_{\mathrm{m}}$, where we utilize dimensionsless density parameters $\Omega_{\mathrm{i}}=\rho_{\mathrm{i}}/\rho_0$ with $\rho_0=3H^2/ 8\pi G$.
We use a model consistent with Wilkinson Microwave Anisotropy Probe (WMAP) 5-year data \cite{Hinshaw:09}. 
The potential \eqref{SteinerPot} and therefore the cosmic evolution is governed by two very different energy scales: the huge Planck mass $m_\mathrm{{pl}}$ and the much smaller energy density $\rho_0$. Explicitly, one derives from \eqref{SteinerPot} for $\phi\rightarrow0$
\be
\lb{0}
V_{\mathrm{Q}}(\phi)\sim\frac{1}{\phi^2} 
\ee
and, respectively, for $\phi\rightarrow\infty$
\be
\lb{infty}
V_{\mathrm{Q}}(\phi)\sim\exp\,(-2B\phi) \ , 
\ee
which is in accordance with the general behavior of a quintessence potential.

\begin{figure}[!ht]
 \begin{center}
 \includegraphics[width=7cm]{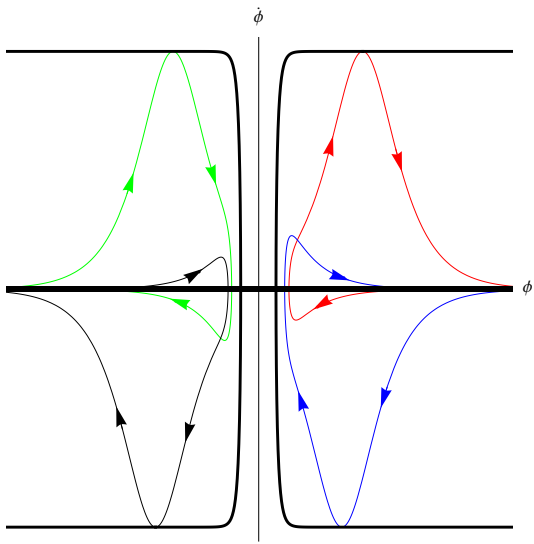}
\caption{Phase portrait with the quintessence potential \eqref{SteinerPot}. The outer boundary (thick line) corresponds to $\rho=\rho_{\mathrm{crit}}$, the thin lines show the solutions of \eqref{phiddot} for different initial data.}\label{fig:Steiner_paramplot}
\end{center}
\end{figure}

\begin{figure}[!ht]
 \begin{center}
 \includegraphics[width=7.5cm]{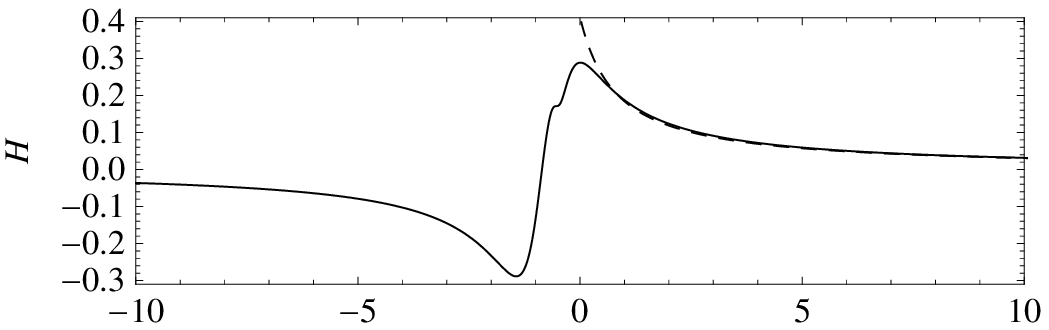}
\includegraphics[width=7.5cm]{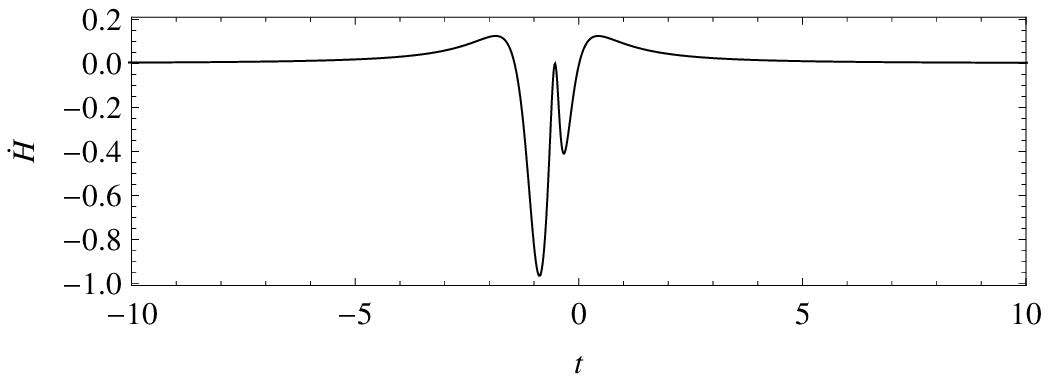}
\caption{Behavior of $H$ and $\dot{H}$ with the quintessence model potential \eqref{SteinerPot}. The dashed line shows the classical solution, whereas the solid lines show the solutions obtained from LQC.}\label{fig:Steiner_cl_vs_LQC}
\end{center}
\end{figure}

\begin{figure}[!ht]
 \begin{center}
 \includegraphics[width=7.5cm]{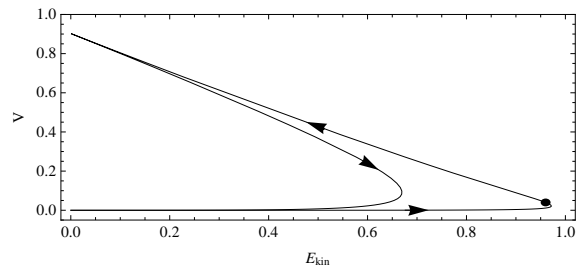}
\caption{Phase portrait of the kinetic energy $\dot{\phi}^2/2$ and the potential energy $V_{\mathrm{Q}}(\phi)$, where the dot corresponds to the bounce. The kinetic energy vanishes at $t=0.5$.}\label{fig:Steiner_kinV}
\end{center}
\end{figure}

Numerical solutions for the time evolution in LQC with a quintessence potential are shown in Fig.~\ref{fig:Steiner_cl_vs_LQC}, respectively, Fig.~\ref{fig:Steiner_kinV}, and the phase potrait, consisting ($\phi, \dot{\phi}$), for different initial values is presented in Fig.~\ref{fig:Steiner_paramplot}. As in the previous cases both the Hubble parameter $H$ and the energy density $\rho$ are bounded and subject to the constraints \eqref{HB}. As can be seen from Fig.~\ref{fig:Steiner_cl_vs_LQC} the Hubble rate starts with a small negative value. For times far away from the bounce the potential energy is the dominant contribution, as can be seen in Fig.~\ref{fig:Steiner_kinV}. All trajectories start from $\phi\rightarrow \pm\infty$ for $t\rightarrow-\infty$, cf. Fig.~\ref{fig:Steiner_paramplot}. When the kinetic energy reaches a value such that the energy density becomes comparable to $\rho_{\mathrm{crit}}$ the magnitude of the Hubble rate starts increasing and quickly becomes zero at $\rho=\rho_{\mathrm{crit}}$ or equivalently at $t=-0.86$. As shown in Fig.~\ref{fig:Steiner_paramplot} the evolution than reaches a point at the outer boundary, where the universe bounces. The bounce implies a change of the sign of the Hubble rate, cf. Fig.~\ref{fig:Steiner_cl_vs_LQC}. Immediately after the bounce the universe expands quickly and reaches a plateau at time $t=-0.5$, where the potential energy reaches a global maximum, whereas the kinetic energy $\dot{\phi}/2$ vanishes. As presented in Fig.~\ref{fig:Steiner_cl_vs_LQC}, for $t\rightarrow\infty$ the Hubble parameter decreases at a radiationlike rate, i.e. $H\sim(2t)^{-1}$ and all trajectories starting from $\phi\rightarrow -\infty$ for $t\rightarrow-\infty$ go back to $\phi\rightarrow -\infty$ and, respectively, all trajecories coming from  $\phi\rightarrow \infty$ end in $\phi\rightarrow \infty$ for large times. Thus, there exist two independent sectors in the phase diagram, cf. Fig.~\ref{fig:Steiner_paramplot}. 

\section{Discussion and Conlusions}\label{sec:conclusion}
As an attempt to solve the shortcomings of the concordance model of cosmology several models of scalar fields have been proposed which interpolate between two stages of the evolution of our universe. We studied the influence of three types of scalar fields of cosmological interest, nameley the Chaplygin gas, a modificated version of it and quintessence, and one more exotic type called the anti-Chaplygin gas. While the first type models a unification of dark matter and dark energy, the second one interpolates between an early inflationary phase and radiation. Contrary to quintessence which was introduced as an effort to describe dark energy in terms of a scalar field, the anti-Chaplygin can be considered as a toy model without any direct application to cosmology.

We presented the solutions to the Friedmann equations in LQC for these four models. We showed that the evolution of the first three models (Chaplygin gas, modified Chaplygin gas and quintessence) follows the classical path until it approaches the critical density, where the modification to the Friemann equation gains in importance. As this modification is negative definite the evolution bounces and the Hubble rate changes sign. Some time after the bounce the evolution follows once again the classical trajectory. We showed that, while all the origin in the $(\phi,\dot\phi)$-phase diagram acts as an attractor for the Chaplygin gas, the solutions of the modified version converge toward $\dot\phi\rightarrow0$ and $\phi\rightarrow\pm\infty$. Quintessence behaves in a similar way, except that there are two independent sectors in the $(\phi,\dot\phi)$-phase diagram such that trajectories with positive respectively negative initial $\phi$ always stay positive respectively negative. The situation is radiacally different for the anti-Chaplygin gas where every trajectory starts and ends with a Type II singularity. Depending on the initial data the evolution may go through a bounce, however LQC is, as expected, not able to remove these Type II singularities. Because of this very fact there are also two independent sectors for $\phi$.
\section*{ACKNOWLEDGMENTS}
We would like to thank Mariam Bouhmadi Lopez, Claus Kiefer and Frank Steiner for discussions.

\end{document}